\documentclass[12pt]{article}

\newenvironment{numberedlist}
{\begin{list}{\makebox[20pt]{\hss(\arabic{itemno})\enspace}}
             {\usecounter{itemno}\labelwidth 20pt}}{\end{list}}

\newcounter{itemno}

\newcounter{itemno1}

\newcounter{itemno2}

\newcounter{exno}

\newcounter{defno}

\newenvironment{defn}{\refstepcounter{defno}\medskip \noindent {\bf
Definition \thedefno.\ }}{\medskip}

\newcommand{\sep}{\;\vert\;}

\newcommand{\oprove}{\vdash\kern-.6em\lower.7ex\hbox{$\scriptstyle O$}\,}

\newcommand{\Dscr}{{\cal D}}
\newcommand{\Pscr}{{\cal P}}

\newcommand{\pderivation}{{\cal P}\kern -.1em\hbox{\rm -derivation}}
\newcommand{\pderivationl}{{\cal P}\kern -.1em\hbox{\em -derivation}}
\newcommand{\pderivable}{{\cal P}\kern -.1em\hbox{\rm -derivable}}
\newcommand{\pderivablel}{{\cal P}\kern -.1em\hbox{\em -derivable}}
\newcommand{\pderivations}{{\cal P}\kern -.1em\hbox{\rm -derivations}}
\newcommand{\pderivability}{{\cal P}\kern -.1em\hbox{\rm -derivability}}

\newcommand{\all}{\forall}
\newcommand{\some}{\exists}

\newcommand{\ie}{{\em i.e.}}

\newsavebox{\lpartfig}
\newsavebox{\rpartfig}

\newenvironment{exmple}{
 \begingroup \begin{tabbing} \hspace{2em}\= \hspace{3em}\= \hspace{3em}\=
\hspace{3em}\= \hspace{3em}\= \hspace{3em}\= \kill}{
 \end{tabbing}\endgroup}

\newcommand{\lb}{\langle}
\newcommand{\rb}{\rangle}
\newcommand{\pr}{prov}

\newcommand{\prove}{exec} 
  
     {\\* \hspace*{\fill} \end{trivlist}}

\newcommand{\muprolog}{{Prolog$^{Ind,G}$}}

\renewcommand{\pr}{pv}
\renewcommand{\psi}{\Sigma}
\newcommand{\df}{\Delta}
\newcommand{\dfm}{\Delta'}
\renewcommand{\prove}{ex} 
\newcommand{\defeq}{\mathrel{\stackrel{{\scriptstyle\triangle}}{=}}}
\renewcommand{\Pscr}{{\cal G}}
\newcommand{\Ascr}{{\cal A}}

\begin{document}

\begin{center}
{\Large {\bf Incorporating   Inductions and Game Semantics
into Logic Programming}}
\\[20pt] 
{\bf Keehang Kwon}\\
Dept. of Computer  Engineering, DongA University \\
Busan 604-714, Korea\\
  khkwon@dau.ac.kr\\
\end{center}

\noindent {\bf Abstract}: 
Inductions and game semantics are two useful extensions to
traditional logic programming.  To be specific, inductions  can capture
a wider class of provable formulas in logic programming.
Adopting game semantics can make logic programming more interactive.

  In this paper, we propose an execution model for a logic language with these  features.
This execution model follows closely the reasoning process in real life.

{\bf keywords:} induction, game semantics, read, computability logic.

\section{Introduction}\label{sec:intro}

Fixed-point definitions, inductions and game semantics are all useful extensions to the theory of
logic programming.  In this paper, we propose an execution model that combines these three concepts.

First, logic programming with fixed-point definitions has been studied by several researchers
 \cite{mcdowell97lics,tiu05eshol}. In this setting, 
clauses  of the form  $A\defeq B$ -- called {\em definition} clauses -- are used to provide
least fixed-point definitions of atoms. We assume that a set $\Dscr$ of such definition 
clauses -- which we call a program -- has been fixed.   The following {\em
  definition-right} rule, which is a variant of the one used in 
LINC\cite{tiu05eshol}, is used in this paper
as an inference  rule which introduces atomic
formulas on the right. \\


$\pr(\sigma,\Pscr\vdash A)$ if $A'\defeq B\in\Dscr$ and  $A'\theta=A\sigma$ and
$\pr(\sigma\theta,\Pscr\vdash B)$. \\

This rule is similar to backchaining in Prolog with the difference that a current answer
 subsititution $\sigma$ (also called a {\em run}) is maintained and 
applied to formulas in a lazy way here.
 The {\em definition-left} rule represents a case analysis in reasoning.\\


$\pr(\sigma, A:\Pscr\vdash D)$ if, for each $\theta$ which is the $mgu(A\sigma, A'
)$ for some $A'\defeq B\in\Dscr$,
    $\pr(\sigma\theta, B:\Pscr\vdash D)$. \\

\noindent Here, $D$ represents a goal.
This rule is well-known and used to instantiate the  free variables of
the sequent by
$\theta$, which is a most general unifier (mgu) for
atoms $A\sigma$ and $A'$.   If there is no such $\theta$, the sequent is proved.

Natural number induction is also useful in many applications.
We use $0$ for zero and $x+1$ for a successor of $x$.
  The following {\em
  nat-right} rules  introduce natural numbers  on the right. \\

$\pr(\sigma,\Pscr\vdash nat(z))$. \\

$\pr(\sigma,\Pscr\vdash nat(I+1))$ if 
$\pr(\sigma,\Pscr\vdash nat(I))$. \\

 The {\em nat-left} rule corresponds to  an induction  in reasoning.\\



  $\pr(\sigma, nat(n)\vdash G)$ if  $\pr(\sigma,\emptyset \vdash G(n/0))$ and
$\pr(\sigma, G(n/j) \vdash  G(n/(j+1)))$  where $j$ is a new variable.

This rule is a well-known induction rule \cite{mcdowell97lics} and used to prove a goal $G$ for all natural 
numbers using only trivial 
inductions. As we shall see later, even simple inductions make their implementation difficult.

The operational semantics of these languages \cite{mcdowell97lics} 
is typically based on intuitionistic provability.
In the operational semantics based on  provability,
solving the universally quantified goal $\all x D$ from a definition $\Dscr$
 simply {\it terminates} with a success if it is provable.

In this paper, we  make the above  operational semantics more  ``interactive'' 
by adopting the game semantics in \cite{Jap03,Jap08}.
That is, our approach in this paper involves a modification of the  operational
semantics to allow for more active participation from the user.
Solving  $\all x D$ from a program $\Dscr$ now  has the
following two-step operational semantics:

\begin{itemize}

\item Step (1): the machine tries to prove $\all x D$ from a program $\Dscr$. If it fails, the
machine returns the failure. If it succeeds, goto Step (2).

\item Step (2): the machine requests the user to choose a constant $c$ for $x$ and then proceeds
with solving the  goal, $[c/x] D$.

\end{itemize}

As an   
illustration of this approach, let us consider the following program. \\

$ \{\ fact(0,1) \defeq \top.\ $

$   fact(X+1,XY+Y) \defeq  fact(X,Y)\ \} $ \\

\noindent 
 As a particular example, consider a goal task 
 $\all x (nat(x) \supset \exists y  fact(x,y))$.

  To prove that this goal is valid, we need to use induction.
Most theorem provers  simply terminates
with a success as it is solvable.  However, in our 
context,
  execution requires more. To be specific, execution proceeds as follows: the system 
 requests the user to select a particular number for $x$. 
After the  number  -- 
 say, $5$ -- is selected, the system returns $y = 120$.
     As seen from the example above, universally quantified goals in intuitionistic logic
can be used to model the $read$ predicate in Prolog.

In this paper we present the syntax and semantics of this  language called \muprolog. 
The remainder of this paper is structured as follows. We describe \muprolog\
   in
the next section.  Section 3 describes the new semantics.
Section 4 concludes the paper. 

\section{An Overview of \muprolog}

Our language is  a variant  of the level 0/1 prover in \cite{tiu05eshol} extended with simple inductions.
 Therefore, we closely follow their presentation in 
\cite{tiu05eshol}.
 We assume that  a program -- a set of definition clauses $\Dscr$ --
is given. 
We have two kinds of goals given  
by $G$- and $D$-formulas  below:
\begin{exmple}
\>$G ::=$ \>  $\top \sep \bot \sep nat(x) \sep A \sep   G \land  G \sep   \some x\ G   $ \\   \\
\>$D ::=$ \>  $\top \sep \bot \sep nat(x) \sep A  \sep D \land D \sep    \some x\ D  \sep \all x\ D \sep  G \supset D \sep nat(x) \supset G$\\
\end{exmple}
\noindent
In the rules above, $A$  represents an atomic formula.

The formulas in this languages are divided into {\em level-0} goals, 
given by $G$ above, and {\em level-1} goals, given by $D$. 
We assume that atoms are partitioned
 level-0 atoms and level-1 atoms. Goal formulas can be level-0 
or level-1 formulas, and in a definition $A \defeq B$, $A$ and $B$ can
be level-0 or level-1 formulas, provided that  level($A$)
$\geq$   level($B$). 

Proving Level-0 formulas and Level-1 formulas is similar to
proving goal formulas in Prolog.
However, there are some major differences:

\begin{itemize}

\item when the Level-1 prover meets the implication $G \supset D$ where $G$ is not $nat(x)$, it
attempts to solve $G$ (in level-0 mode). If $G$ is solvable with  all the possible  answer
substitutions $\psi_1,\ldots,\psi_n$, then the Level-1 prover  checks that, for every 
substitution $\psi_i$, $D\psi_i$ holds.
If Level-0 finitely fails, the implication is proved. 

\item when the Level-1 prover meets the implication $nat(x) \supset G$, the choices for $x$ can be infinite.
Therefore the machine needs to prove $G$ using induction (in induction mode). In induction mode,
the machine attempts to decompose the induction hypothesis $G(x/n)$ (in level-0 submode) into a set atomic formulas $\Ascr$. Then it attempts to solve $G(x/n+1)$ (in level-1 submode
) relative to $\Ascr$.
If $G(x/n+1)$ is solvable with respect to $G(x/n)$ with  an (partial) answer
substitution $\df_n$ , then the machine concludes that  $G(x/k)$
 holds with an (total) answer substitution $\df_k\ldots\df_0$ (\ie, by composing answer substitutions)
for each natural number $k$.

\end{itemize}

 We will  present the standard operational 
semantics for this language  as inference rules \cite{Khan87}. 
 Below the notation $G:\Pscr$ denotes
$\{ G \}\cup\Pscr$. Note that execution  alternates between 
two phases: the left rules phase
and the right rules phase.
In this fragment, all the left rules (excluding the defL in in) are 
invertible and therefore  the left-rules (excluding the defL) take precedence over the right rules.
Note that our semantics is a lazy version of the semantics of level 0/1 prover in the sense that
an answer substitution is applied as lazily as possible. 
Below, the proof procedure for some formula returns a final run $\Sigma$  in normal mode and a final run 
$\Delta$ in induction mode. Note that it is not always possible to obtain the final run due to the
presence of induction. In such a case, we assume that the machine returns a $Failure$.

\begin{defn}\label{def:semantics}
Let $\sigma, \delta$ be answer substitutions, let $G, D$ be a goal, let $\Pscr$ be a set of
$G$-formulas. 
Then the task of 

\begin{itemize}

\item proving $D$ from $\emptyset$ (empty premise) with respect to $\sigma,\Dscr$ and returns  a total
run $\psi$ -- $\pr(l_1,\sigma,\emptyset,D,\psi)$ -- \% in level 1,

\item proving $D$ from $G:\Pscr$ with respect to $\sigma,\Dscr$ and returns a total run
$\psi$ -- $\pr(l_0,\sigma,G:\Pscr,D,\psi)$ -- \% in level 0,

\item proving $G$ from $G:\Pscr$ with respect to $\sigma,\delta,\Dscr$ and returns a partial run 
$\df$ -- $\pr(i_0,\sigma,\delta,G:\Pscr,G,\df)$ --  \% induction mode, level 0

\item proving $G$ from $G:\Pscr$ with respect to $\sigma,\delta,\Dscr$ and returns a partial run
$\df$ -- $\pr(i_1,\sigma,\delta,G:\Pscr,G,\df)$  \%  induction mode, level 1
\end{itemize}

-- are defined as 
follows:

\begin{numberedlist}

\item  $\pr(l_0,\sigma, \bot:\Pscr \vdash D,\sigma)$. \% This is a success.

\item    $\pr(l_0,\sigma, \top:\Pscr \vdash D,\psi)$ if  
 $\pr(l_0,\sigma,\Pscr \vdash D,\psi)$. \% $\top$ in the premise is redundant.

\item $\pr(l_0,\sigma,A:\Pscr \vdash D\theta,\psi)$ if, for each $\theta$ which is the $mgu(A\sigma, A')$ for some $A'\defeq B\in\Dscr$,
    $\pr(l_0,\sigma\theta, B:\Pscr \vdash  D,\psi)$. \% DefL rule

\item $\pr(l_0,\sigma, nat(n):\Pscr \vdash G, Failure)$ if \% invokes induction \\
\hspace{4em} $\pr(l_1,\sigma\{(n,0) \}, \emptyset \vdash
 G,\psi)$  \% prove base case \\
\hspace{4em} and  \\
\hspace{4em} $\pr(i_0,\sigma\{(n,j) \},\emptyset, G \vdash G(n/n+1),\df)$ \% prove induction step \\
 where  $j$ is a  new free 
variable. \% In induction step,  $\delta$ -- a partial substitution --  is initialized to an empty substitution. $Failure$ means that it is not possible to obtain the final run.

\item    $\pr(l_0,\sigma, (G_0 \land G_1):\Pscr\vdash D,\psi)$ if   $\pr(l_0,\sigma, G_0:G_1:\Pscr \vdash D,\psi)$. 



\item    $\pr(l_0,\sigma, \some x G:\Pscr \vdash D,\psi)$ if   $\pr(l_0,\sigma, [y/x]G:
\Pscr \vdash  D,\psi)$ where $y$ is a $new$ free variable.

\% Below is the description of the   level-0 prover in induction phase 

\item $\pr(i_0,\sigma,\delta,\Ascr\vdash G,\df)$  if   $\pr(i_1,\sigma,\delta, \Ascr\vdash G,\df)$. \% switch from $i_0$ to $i_1$.

\item $\pr(i_0,\sigma,\delta, A:\Pscr \vdash G,\df)$  if (a nonatomic $G$ is in $\Pscr$) and  
$\pr(i_0,\sigma,\delta, G:A:\Pscr' \vdash G,\df)$. 
where $\Pscr'$ is $\Pscr-G$.   \% process $\Pscr$ if it contains a nonatomic formula.

\item    $\pr(i_0,\sigma,\delta,(G_0 \land G_1):\Pscr \vdash G,\df)$ if   $\pr(i_0,\sigma,\delta, G_0:G_1:\Pscr\vdash  G,\df)$. 



\item    $\pr(i_0,\sigma,\delta,\some x G_1:\Pscr\vdash G, \df)$ if   $\pr(i_0,\sigma,\delta, [y/x]G_1:
\Pscr\vdash G,\df)$ where $y$ is a $new$ free variable. 

\% Below is the description of the   level-1 prover in induction phase 

\item $\pr(i_1,\sigma,\delta, A:\Ascr \vdash A,\delta)$. \% This is a success via induction hypothesis $A$.

\item $\pr(i_1,\sigma,\delta, \Ascr\vdash G_0 \land G_1, \dfm_0 \cup \dfm_1)$  if $\pr(i_1,\sigma,
\delta,\Ascr \vdash G_0,\df_0)$ and 
$\pr(i_1,\sigma,\delta, \Ascr\vdash G_1,\df_1)$. 

Here, the answer substitution $\dfm_0$ is  identical to $\df_0$ but  locations of the form $loc(x)$
 in $\dfm_0$ are  adjusted to new locations
properly. Similarly for $\dfm_1$.

\item $\pr(i_1,\sigma,\delta,\Ascr\vdash \some x G,\df)$  if $\pr(i_1,\sigma,\delta\delta_1,\Ascr \vdash  [y/x]G,\df)$ 
where $y$ is a new free variable, $\delta_1 =  \{ (loc(x),t) \} \{ (y,t) \}$ and $t$ is a term.
Note that we assume that $loc(x)$ represents a unique location in the sequent.

\% Below is the description of the   level-1 prover

\item    $\pr(l_1,\sigma,\emptyset\vdash \top,\sigma)$. \% solving a true goal

\item $\pr(l_1,\sigma,\emptyset\vdash A,\psi)$ if $A'\defeq B\in\Dscr$ and  $A'\theta = A\sigma$ and
$\pr(l_1,\sigma\theta,\emptyset\vdash B,\psi)$. \% DefR

\item $\pr(l_1,\sigma,\emptyset\vdash D_0 \land D_1,\psi_0 \cup \psi_1)$  if $\pr(l_1,\sigma,\Pscr \vdash D_0,\psi_0)$ and 
$\pr(l_1,\sigma, \Pscr \vdash D_1,\psi_1)$.  \% conjunctive goals


 \item $\pr(l_1,\sigma,\emptyset \vdash G \supset D,\psi)$  if $\pr(l_0,\sigma,G \vdash D,\psi)$. \% switch from level 1 to level 0

 \item $\pr(l_1,\sigma,\emptyset \vdash \all x D,\psi)$  if $\pr(l_1,\sigma,\emptyset \vdash
 [y/x]D,\psi)$ where $y$ is a $new$ free variable.

\item $\pr(l_1,\sigma,\emptyset \vdash \some x D,\psi)$  if $\pr(l_1,\sigma\sigma_1,\emptyset \vdash [y/x]G,\psi)$ 
where $y$ is a new free variable, $\sigma_1 =  \{ ( y,t) \}$ and $t$ is a term.



\end{numberedlist}
\end{defn}

The following is a proof tree (from bottom up) of the example given in Section 1.
Note that a proof tree is represented  as a list.
Now,  a proof tree of a {\it proof formula} 
 is a list of tuples of
the form $\lb E,\psi,Ch \rb$ where $E$ is a proof formula, $\psi$ is a final run for $E$, and $Ch$ is  a list
of the form $i_1::\ldots::i_n::nil$ where each $i_k$
is the address of
its $k$th child (actually the distance to $E$'s $k$th chilren
in the proof tree). 

\% base case \\

$l_1$,$\{ (h_0,0),(w_0,1) \}$, $\emptyset\vdash \top$,$\Sigma$, nil \% success

$l_1$,$\{ (h_0,0),(w_0,1) \}$, $\emptyset\vdash fact(h_0,w_0)$, $\Sigma$, 1::nil \% defR

$l_1$,$\{ (h_0,0) \}$, $\emptyset \vdash\some y\ fib(h_0,y)$, $\Sigma$, 1::nil \% $nat$-0 \\ \\

\% start of induction step 

$i_1$,$\{ (h_0,j) \}$, $\{ (y,w_0),(loc(z),(j+1)w_0),(w_1,(j+1)w_0) \}$, $  fact(h_0,w_0) \vdash 
fact(h_0,w_0)$, $\Delta$, nil \% success

$i_1$,$\{ (h_0,j) \}$, $\{ (y,w_0),(loc(z),(j+1)w_0),(w_1,(j+1)w_0) \}$, $ fact(h_0,w_0) 
\vdash fact(h_0+1,w_1)$,  $\Delta$, 1::nil \% defR

$i_0$,$\{ (h_0,j) \}$, $\{ (y,w_0) \}$, $fact(h_0,w_0)\vdash \some z fact(h_0+1,z)$, $\Delta$, 1::nil
 \% $\some$-L

$i_0$,$\{ (h_0,j) \}$,$\emptyset$, $ \some y fact(h_0,y)\vdash  \some z fact(h_0+1,z)$, $\Delta$, 1::nil \% $\some$-L

\% end of induction step \\

$l_0$,$\emptyset$, $nat(h_0) \vdash \some y\ fib(h_0,y)$, $Failure$, 5::1::nil \% defL

$l_1$,$\emptyset$, $\emptyset \vdash nat(h_0) \supset \some y\ fib(h_0,y)$, $Failure$, 1::nil

$l_1$,$\emptyset$,$\emptyset \vdash  \all x  ( nat(x) \supset \some y\ fib(x,y))$, $Failure$, 1::nil \% $\all$-R \\

\noindent In the above, $\Sigma = \{ (h_0,0),(w_0,1) \}$ and $\Delta = \{ (y,w_0),(loc(z),(j+1)w_0),(w_1,(j+1)w_0) \}$.
\section{An Alternative Operational Semantics}\label{sec:0627}

Adding game semantics  requires some changes to the previous execution model.
To be precise, our new execution model -- adapted from \cite{Jap03} -- solves the goal relative to the program using the proof tree built in the proof search.

To be precise,  execution  proceeds in  two different phases: normal phase
and induction phase.
In normal phase, execution simply follows the proof tree because the proof tree encodes all the possible
total runs.  In induction phase, things are more complicated. Note that the proof tree in induction mode
encodes only the partial run (from
$i$th inductive step to $i+1$th inductive step).
Therefore, a total run must be  obtained from composing all the partial runs, not from
the proof tree.

 In addition, to deal with the universally quantified goals properly,
the execution needs to maintain an $input$ $substitution$ $F$ of the form
$\{ y_0/c_0,\ldots,y_n/c_n \}$ where each $y_i$ is a variable introduced by a universally quantified goal
in the proof phase and each $c_i$ is a user input  during the execution phase.

\begin{defn}\label{def:exec}
 let $L$ be a fixed proof tree.
Let $i$ be an index to a proof tree and let  $F$ be an input substitution.
In addition, let $\sigma$ be an answer substitution, let $\Delta$ be an  answer substitution (obtained
from composing induction steps). 
Then  executing $L_i$ (the $i$ element in $L$) with $F$ in normal phase -- written as $\prove(i,F)$ --
and executing $G$ with $\sigma,\Delta,F$ in induction phase -- written as $\prove(ind,\sigma,
\Delta,\emptyset\vdash G,F)$ --
are defined as follows: 

\begin{numberedlist}

\item  $\prove(i,F)$ if $L_i = (E,nil)$. \% no child. This is a success.

\item  $\prove(i,F)$ if $L_i = ((l_1,\sigma,\emptyset,D_0 \land D_1,\psi), m::1::nil)$ and \\
 $\prove(i-m,F)$ and \% execute $D_0$ \\
$\prove(i-1,F)$. \% execute $D_1$

\item  $\prove(i,F)$ if $L_i = ((l_1,\sigma,\Pscr,\all x D,\psi), 1::nil)$ and  \\
$L_{i-1} = ((l_1,\sigma,\Pscr, [y/x] D,\psi), \_)$ and \\
 $read(r)$ \% read a user input \\
and $\prove(i-1, F \cup \{ y/c \})$ \% update $F$ for universal quantifiers \\
 where $c$ is the user input (the value 
stored in $r$).

\item  $\prove(i,F)$ if $L_i = ((l_0,\sigma,A:\Pscr, D,\psi), i_1::\ldots::i_n::nil)$ and \\
choose a $i_k$ such that $L_{i-i_k} = ((l_0,\sigma\theta_k,B,\Pscr, D,\psi), \_)$ and \\
($F$ and $\theta_k$ agree on the variables appearing in $F$) \\
  and $\prove(i-i_k, F)$. \% choose a correct one using $F$ among many paths in defL
 
\item  $\prove(i,F)$ if $L_i = ((l_0,\sigma, nat(n) \vdash G,Failure),  p::q::nil)$ and \\
 $L_{i-p} = ((l_1,\sigma \{ (n,0) \}, \emptyset \vdash G, \psi_B), \_)$ and  \% base case \\
$L_{i-q} = ((i_0,\sigma \{(n,j \}, \emptyset, G \vdash G(n/n+1), \Delta),  \_)$ and \% induction step \\
$\prove(ind,\sigma,\Delta_{total},\emptyset\vdash G, F)$  \% run in induction mode \\
where $k = F(n)$ \\
and $\Delta_{total} = (\df|(j,k-1)\ \ldots\ \df|(j,0) \psi_B)) | (j,k-1) $ represents a total run for $G$


\item  $\prove(i,F)$ if $L_i = ((l_1,\sigma, \emptyset\vdash \some x D), 1::nil)$ and \\
$L_{i-1} = ((l_1,\sigma\{ (y,t) \}, \emptyset \vdash [y/x] D), \_)$ \\
 and (print $x = y \sigma F$) and $\prove(i-1,F)$. 
 Hence the value of $x$ is $y$ instantiated by $F$ and $\sigma$.

\item   $\prove(i,F)$ if $L_i = (E,1::nil)$  and $\prove(i-1,F)$. \% otherwise

\item  $\prove(ind,\sigma,\Delta,\emptyset\vdash A,F)$. \% success in induction mode

\item  $\prove(ind,\sigma,\Delta,\emptyset\vdash G_0\land G_1,F)$ if \\
$\prove(ind,\sigma,\Delta,\emptyset\vdash G_0,F)$ and \% execute the first goal. \\
$\prove(ind,\sigma,\Delta,\emptyset\vdash G_1,F)$. \% execute the second goal. 

\item  $\prove(ind,\sigma,\Delta,\emptyset\vdash \some x G,F)$ if 
 (print $x = t$) and \\ 
$\prove(ind,\sigma,\Delta,\emptyset\vdash [t/x]G,F)$ \\
where $t = loc(x)\ \Delta\sigma F$. \% apply $\Delta$, $\sigma$ and then $F$ to $loc(x)$. \\

\end{numberedlist}
\end{defn}

\noindent  
Initially, $\sigma, F$  are empty substitutions.

 In the above, 
$\Delta_{total} = ((\df|(j,k-1) \ldots\ \df|(j,0)\ \psi_B)) | (-k+1) $ is used to correctly
obtain a total run for $G$. To be precise, the notation $\df|(j,i)$ is used  

\begin{itemize}

\item to rename each varaible $w_r$ to $w_{r+im}$,

\item to replace $j$ with $i$ 

\end{itemize}
\noindent where $m$ is the number of existentially quantified variables in $G$.
Thus the composition $\df|(j,k-1)\ \ldots\ \df|(j,0)\ \psi_B$ contains all the answer substitutions obtained in
inductive steps upto the number $k$. Thus it contains all the answer substitutions for $km$ variables. 
Then to produce correct answers in solving $G$, we must
undo the renaming via $| (-k+1)$, deleting unnecessary answer substitions. 
Note that each $\Delta|(j,i)$ may contain location variables of the form $loc(x)$ and we assume that
$loc(x)$ is adjusted properly in obtaining $\Delta_{total}$.

The following is an execution sequence of the goal
$\all x  (nat(x) \supset \some y\ fib(x,y))$ using the  proof tree above. We assume that
 the user chooses $3$ for $x$.
Note that the last  component represents $F$. \\






\% execution (from bottom up) \\




ind, $i_0$,$\emptyset$, $\Delta_{total}$, $fact(h_0+1,6)$, 1::nil \% success, print z = 6.

ind, $i_0$,$\emptyset$, $\Delta_{total}$, $\some z fact(h_0+1,z)$, 1::nil \% $\some$-L \\

$l_0$,$\emptyset$, $nat(h_0)\vdash \some y fib(h_0,y)$,$\_,\_$, 5::1::nil, $\{ (h_0,3) \}$\% defL

$l_1$,$\emptyset$, $\emptyset \vdash nat(h_0) \supset \some y fact(h_0,y)$,$\_,
 \_$, $\{ (h_0,3) \}$\% the user input is 3. update $F$

$l_1$,$\emptyset$,$\emptyset\vdash \all x  ( nat(x) \supset \some y fact(x,y))$, $\_, \_$,
 $\emptyset$ \% $\all$-R \\

\noindent In the above, $\Delta_{total}$ is obtained as follows: 

\begin{numberedlist}

\item From the base case in the proof tree, we obtain $\psi_B = \{ (h_0,0),(w_0,1) \}$.

\item From the inductive case in the proof tree, we obtain $\{ (h_0,j) \}$ and a run  
$\Delta = \{ (loc(z),(j+1)w_0),(w_1,(j+1)w_0)$.

\item Then $\Delta|(j,i) = \{ (loc(z),(i+1)w_i), (w_{i+1},(i+1)w_i) \}$

\item $\Delta|(j,2)\ldots \Delta|(j,0)\ \psi_B = 
 \{ (w_3, (2+1)w_2) \} \{ (w_2, (1+1)w_1) \} \{ (w_1, 1w_0) \} \{ (w_0, 1) \}     = \{ (w_3, 6), (w_2, 2), (w_1, 1), (w_0, 1)  \}$. It also contains answer substitutions for $loc(z_0),\ldots$ which we will not
show here.

\item $\Delta_{total} = (\Delta|(j,2)\ldots \Delta|(j,0)\ \psi_B) | (j,-k+1) = \{ (w_1, 6) \}$

\item In the above, for simplicity, we omit the answer substitutions for $loc(z)$ variables
in $\Delta_{total}$.

\end{numberedlist}

\section{Conclusion}\label{sec:conc}

In this paper, we have considered a new execution model for a subset of 
the level 0/1 prover, enhanced
 with simple inductions
and game semantics.
 This new model is interesting in that it gives a logical status to the $read$ predicate in Prolog.
We plan to connect our execution model to Japaridze's Computability Logic \cite{Jap03,Jap08}
 in the near future.



\bibliographystyle{plain}



\end{document}